\documentclass[cits]{PoS}

\title{The link between ejection of a jet component and
  characteristics of variable continuum emission in the active galaxy 3C\,390.3}

\ShortTitle{Coupling between jet ejection and continuum emission in the radio galaxy 3C\,390.3}

\author{\speaker{Tigran Arshakian}%
  \thanks{On leave from Byurakan
         Astrophysical Observatory, Aragatsotn province, 378433
         Byurakan, Armenia}\\ Max-Planck-Institut f\"ur
         Radioastronomie, Auf dem H\"ugel 69, 53121 Bonn, Germany\\
         E-mail: \email{tigar@mpifr-bonn.mpg.de}}

\author{Tomaso Belloni\\
        INAF-Brera Astronomical Observatory, Via
        E.Bianchi 46, I-23807, Merate, Italy \\
        E-mail: \email{tomaso.belloni@brera.inaf.it}}

\abstract{We study the correlations between the jet ejection event and
  changes in the continuum emission of the radio-loud galaxy
  3C\,390.3, using the archived monitoring data in radio, optical and
  X-ray. We present evidence for the link between the variable optical
  continuum and a stationary radio feature in the jet. The ejection of
  radio components happens during, or after, the dip in the X-ray
  light curve. Moreover, during the X-ray dip the flux variability is
  significantly reduced while the hardness ratio and its variance
  becomes harder. These findings strengthen the idea of similarity
  between active galactic nuclei (AGN) and microquasars, pointing
  towards a common physical mechanism acting in the disk-jet
  system. Other similarities are also discussed on the basis of
  comparision between 3C\,390.3 and the microquasars GRS\,1915+105 and
  Cyg X-1. If the analogy (based on linear mass scaling) between the
  ejection rates of the microquasar GRS\,1915+105 and 3C\,390.3 is
  correct, then the rate of ejections in 3C\,390.3 should vary between
  $\sim$(0.01 to 1) ejections per year on a time scale of thousand
  years.}

\FullConference{VI Microquasar Workshop: Microquasars and Beyond\\
                 September 18-22, 2006\\
                 Como, Italy}

\begin{document}

\section{Introduction}
The AGN are the most powerful objects radiating over the entire
electromagnetic spectrum. The bulk of continuum emission is thought to
be produced by accretion of surrounding material onto the central
massive nucleus (or black hole) which ejects it in the form of a
relativistic collimated outflow (jet) of radio emitting plasma. On
sub-pc-scales, the radio jet exhibits typically stationary and moving
features of enhanced emission detectable with the Very Long Baseline
Interferometry (VLBI) technique \cite{kellermann04}. Moving components
are ejected from the base of the jet very near to the central nucleus
(or black hole). Stationary components of the jet can be associated
with the internal oblique shocks, the emission and geometry of which
depend strongly on the density distribution of the external medium
\cite{gomez95}. The efficiency of coordinated radio VLBI and X-ray
monitoring for investigating the disk-jet couplings was demonstrated
by Marscher et al. \cite{marscher02}. They found that the dips in the
X-ray continuum emission are associated with the epochs of ejection of
radio blobs in the radio-loud galaxy 3C\,120 and interpreted this
correlation as resulting from accretion of some part of the X-ray
emitting disk into the black hole followed by ejection of a new radio
component. Little is known about continuum emission variability and
jet ejection coupling mainly because of the lack of coordinated
multiwavelength monitoring campaigns in radio (VLBI) and other
wavebands. To examine this coupling, we combine the jet ejection
epochs in a radio-loud galaxy 3C\,390.3 ($z$=0.00561) and archived
monitoring data in radio, optical and X-ray ({\it ROSAT} and {\it
RXTE}) bands. The dense X-ray data allows zooming into the regions of
X-ray dips to study the X-ray emission characteristics at the epoch of
a jet ejection event.


\section{Correlations between the sub-pc-scale jet and variability of continuum emission in 3C\,390.3}
\begin{figure}
\includegraphics[angle=-90,width=.6\textwidth]{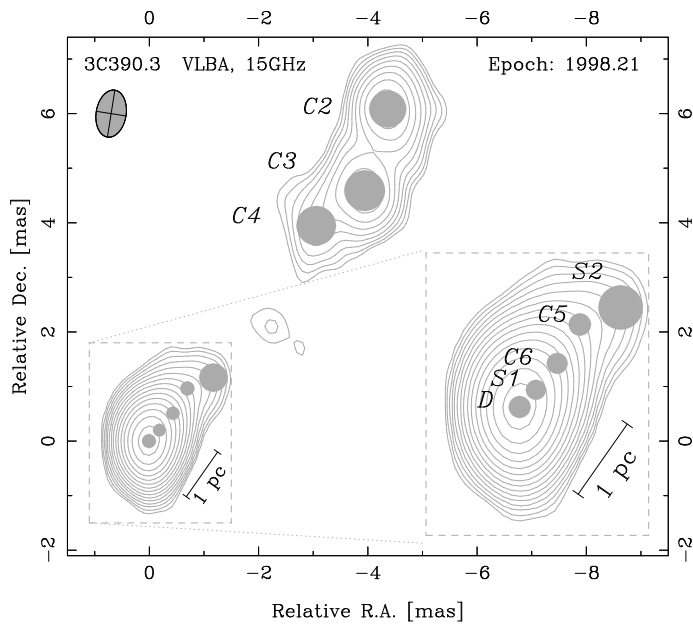}
\caption{VLBA image of 3C\,390.3 made in 1998.21 at 15 GHz
\cite{arshakian05}. Model fit of the jet identified three stationary
(D, S1, S2) and five moving components. The innermost two stationary
components D and S1 are separated within 0.28 mas$\approx$ 0.3 pc of
projected distance. D is identified with the base of the jet and S1 is
likely to be a standing shock of the jet}
\label{fig1}
\end{figure}

Ten epochs of Very Long Baseline Array (VLBA) imaging at 15 GHz
(obtained between 1994 to 2002 as a part of the 2\,cm VLBA survey
\cite{kellermann04}) revealed the structure of the jet, the kinematics
of jet components and the evolution of their emission on angular
scales of $\sim 3$ milliarcseconds (1 mas=1.09 pc)
\cite{arshakian05}. Three stationary features (D, S1 and S2, see
Fig.~\ref{fig1}) and five moving components (C2-C6) with apparent
speeds from 0.8 to 1.5\,{\it c} were identified. The linear fits of
moving components were used to estimate the epoch of their ejection
from the D component (Fig.~\ref{fig2}, red triangles), and the time at
which they pass the stationary feature S1 (gray triangles). It was
found that the bulk of variable radio emission (>90\% on scales of two
milliarcseconds) is radiated predominantly by the two stationary
components D and S1 of the jet separated by $0.28\pm0.03$ mas
($\approx0.3$ pc of projected distance).


\subsection{The link between radio and optical continuum emission}
In Fig.~\ref{fig2}, the long-term variations of radio emission from D,
S1 and the total radio emission (from D and S1) are superimposed onto
optical continuum flux variations at 5100\AA. There is a significant
correlation (>95\%) between variable radio emission from component S1
and optical continuum emission for a time delay of $\sim0.2$\,yr
\cite{arshakian05}, while the total radio flux does not show any
significant correlation with changes in the optical flux. This
indicates that there is a physical link between variable optical
continuum emission and variable radio emission from the stationary
component S1 of the jet, suggesting that the bulk of optical continuum
emission in the radio-loud galaxy 3C\,390.3 is non-thermal and produced in
the jet. Note that the variability amplitudes (see
e.g. \cite{rodriguez97}) of the stationary components S1 and D of the jet
are higher than those of the total radio, optical and X-ray continuum
emission, with the variability amplitude of the S1 component being
higher by a factor of $\ge2$ (Fig.~\ref{fig3}). This is a strong
argument in favor of the fact that the source of variable continuum is localized in the
innermost part of the jet and most likely to be associated with the S1
stationary feature of the jet.

\subsection{The link between the S1 component of the jet and optical continuum emission}
\begin{figure}
\includegraphics[angle=-90,width=.8\textwidth]{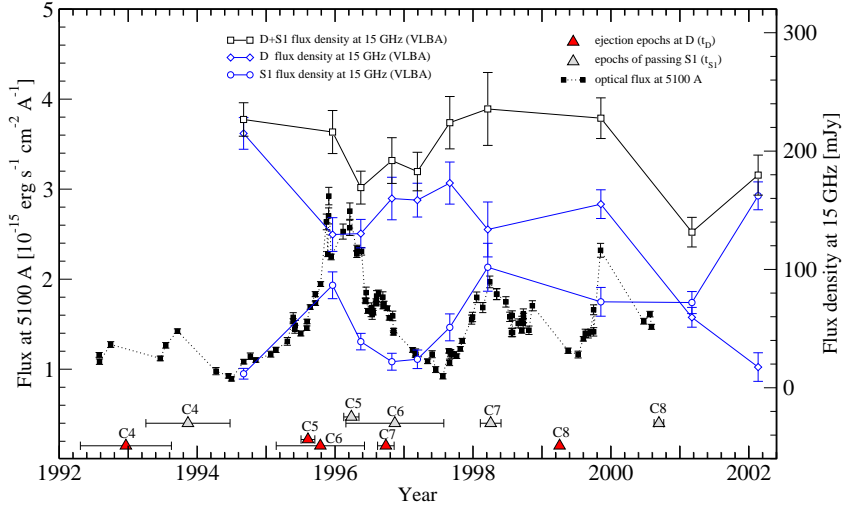}
\caption{The long-term variations of the optical continuum light curve at
  5100\AA\ \cite{shapo01} (black squares) and variations of radio flux
  density from the D, S1 and D+S1 components (empty blue diamonds, circle
  and empty black squares) are superimposed onto the times of ejection
  of radio components from the presumed base of the jet D (red
  triangles) and the times of their separation from the stationary
  component S1 (gray triangles). Adopted from \cite{arshakian05} }
\label{fig2}
\end{figure}

\begin{figure}
\includegraphics[angle=-90,width=.6\textwidth]{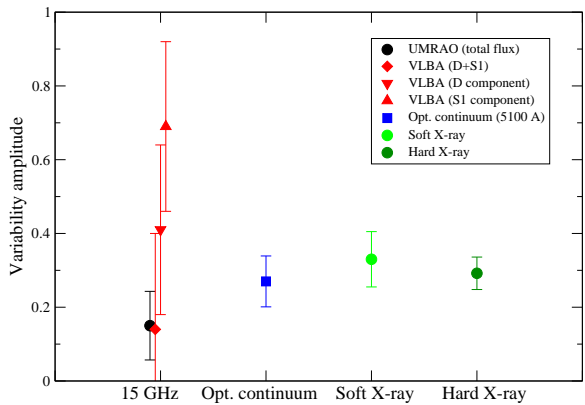}
\caption{Amplitude of variability of continuum emission in 3C\,390.3
  at different frequencies: radio at 15 GHz (single dish total radio
  flux$^{1}$, VLBA flux of D component, S1 and D+S1 components
  \cite{arshakian05}), optical at 5100\,\AA\ \cite{shapo01}, soft
  X-ray at 0.1-2 keV \cite{leighly97a} and hard X-ray at 2-20 keV
  \cite{gliozzi06}.}
\label{fig3}
\end{figure}

Further support for the link between S1 radio component and variable
optical emission is provided by the correlation between the
characteristic time (gray triangles in Fig.~\ref{fig2}) at which the
moving components C4--C7 pass the stationary feature S1. The epochs of
passing of four C4--C7 components at S1 are delayed with respect to
the maxima in the optical continuum (the average time delay is
$0.18\pm 0.06$ years). The null hypothesis that this happens by chance
is rejected at a confidence level of 99.98\% \cite{arshakian05}.

\begin{figure}
\includegraphics[angle=-90,width=.6\textwidth]{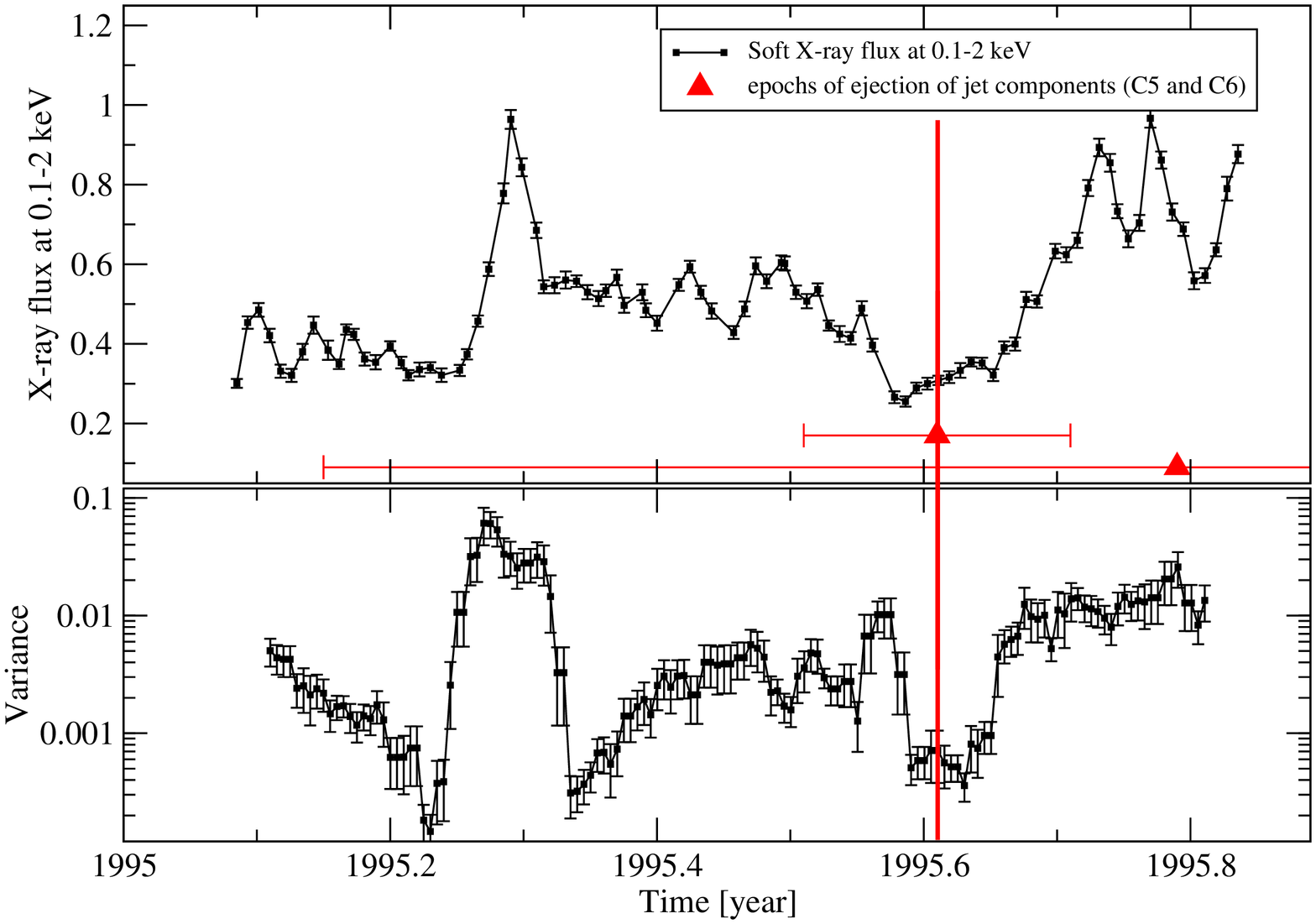}
\caption{ The epochs of ejection of the C5 and C6 radio components (red
  triangles) in 3C\,390.3, time variations of the soft X-ray flux at
  0.1-2 keV (in units of Crab Nebula) and its variance taken over a
  sliding box with a width of 0.05 years.}
\label{fig4}
\end{figure}

\subsection{The link between the D component and characteristics of the X-ray emission}
The D component is likely to be the base of the jet, which is believed
to be located near the central nucleus at a distance of $\ge200$
Schwarzschild radii in or above a hot corona \cite{fabian04}. The
ejection of plasma material from the base of the jet is thought to
be coupled with the accretion of a part of disk material on the central
nucleus. This scenario is supported by the correlation found between
the dip in the X-ray flux associated with the accretion process and
subsequent ejection of the jet component in the radio galaxy 3C\,120
\cite{marscher02}. To check this correlation for 3C\,390.3 and to
explore new correlations, we used the archived X-ray monitoring data
(the dense sampling of X-ray data allows the properties of the X-ray
emission to be studied on time scales from 7-30 days).
\footnotetext[1] {Data are taken from the University of Michigan Radio
Astronomy Observatory online database (H.D.~Aller and M.F.~Aller,
private communication).}

The superposition of the ejection epochs of the C5 and C6 components and the soft
X-ray light-curve (Fig.~\ref{fig4}, upper panel) shows that the
ejections occur during or after the dip in the X-ray flux. An
interesting finding is that the variability of the X-ray flux becomes
significantly lower during the dip in the X-ray flux (Fig.~\ref{fig4},
lower panel) similar to what observed in the
microquasar GRS\,1915+105. The average X-ray
flux is higher by factor 1.5 after the radio ejection in $\sim$1995.6,
and there is a marginal evidence that the flux variability is also
higher after the ejection of the C5 component. This may indicate that
the total X-ray flux is enhanced by beamed variable X-ray emission of
a new born relativistic radio component.

Leighly et al. \cite{leighly97b} pointed out the remarkable feature of
a single flare in 1995.3: the variance of the X-ray flux before and
after the flare is reduced, evidencing that the X-ray variability is
nonlinear. Another interesting feature is the increase of the flux
after the flare by a factor of 2 (Fig.~\ref{fig4}) which may indicate
that the inner radius of the X-ray emitting disk shrinks after the
flare. An interesting question is whether this isolated flare is a
characteristic feature preceding the jet ejection.

\begin{figure}
\includegraphics[angle=-90,width=.52\textwidth]{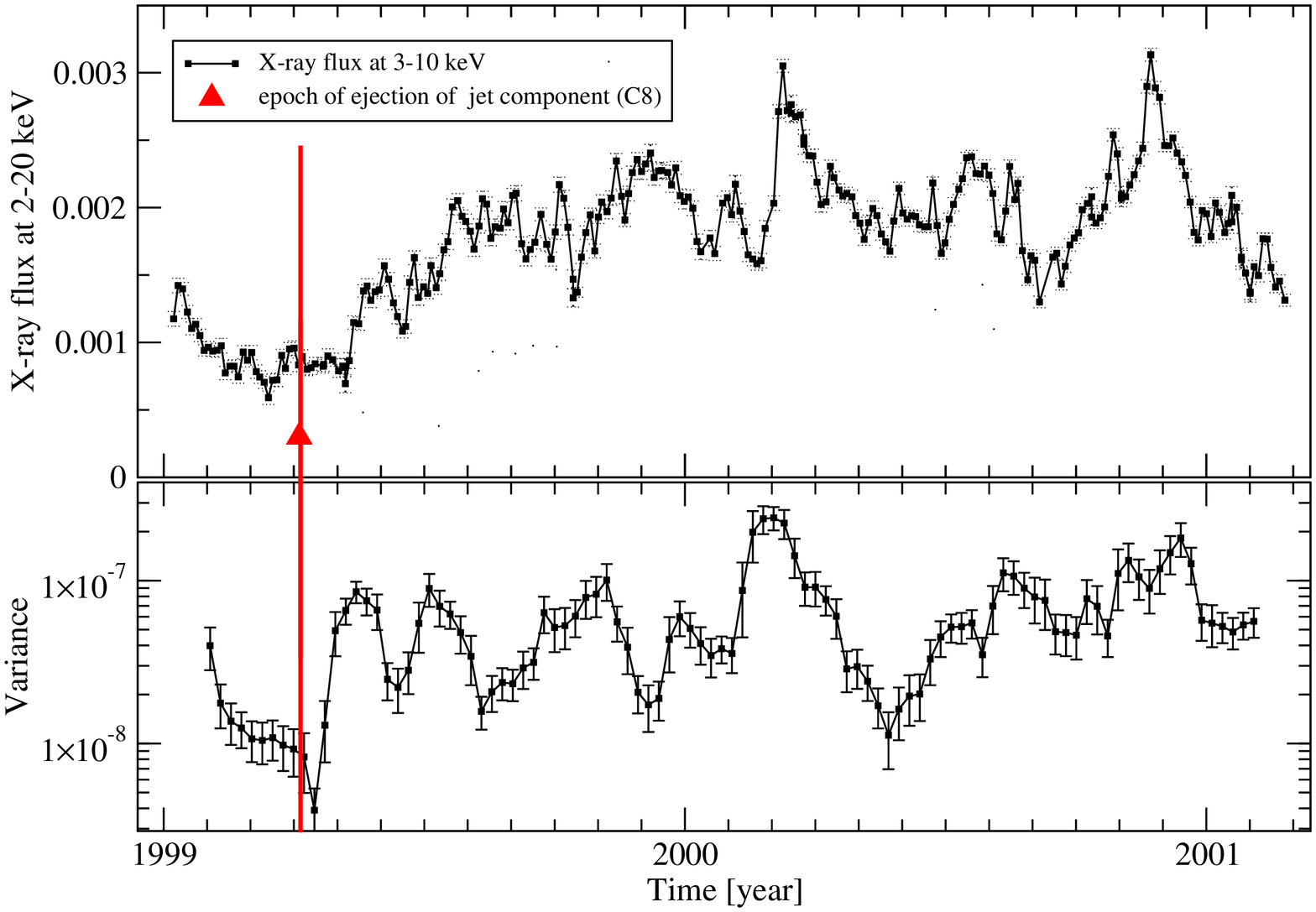}
\includegraphics[angle=-90,width=.52\textwidth]{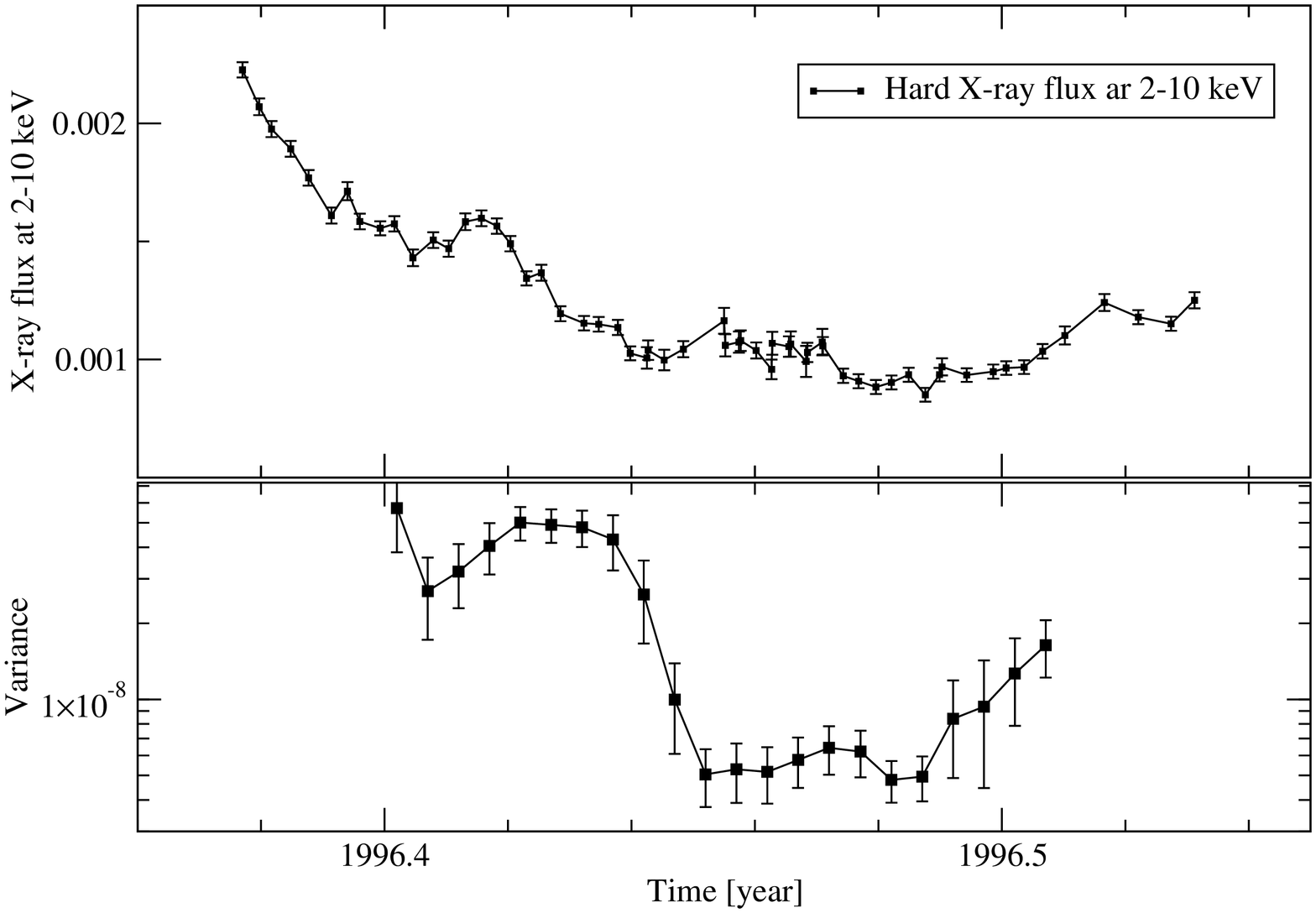}
\includegraphics[angle=-90,width=.52\textwidth]{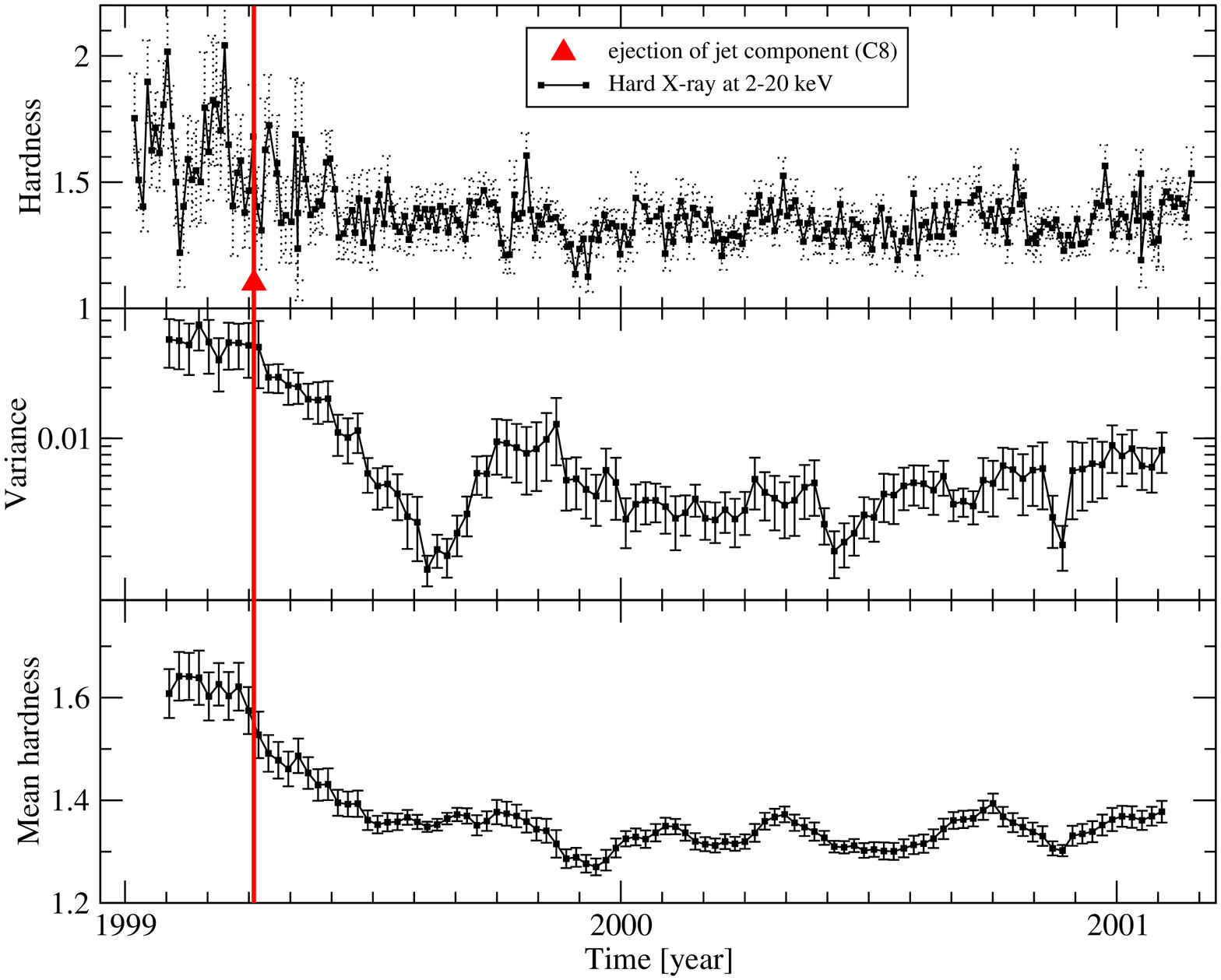}
\includegraphics[angle=-90,width=.52\textwidth]{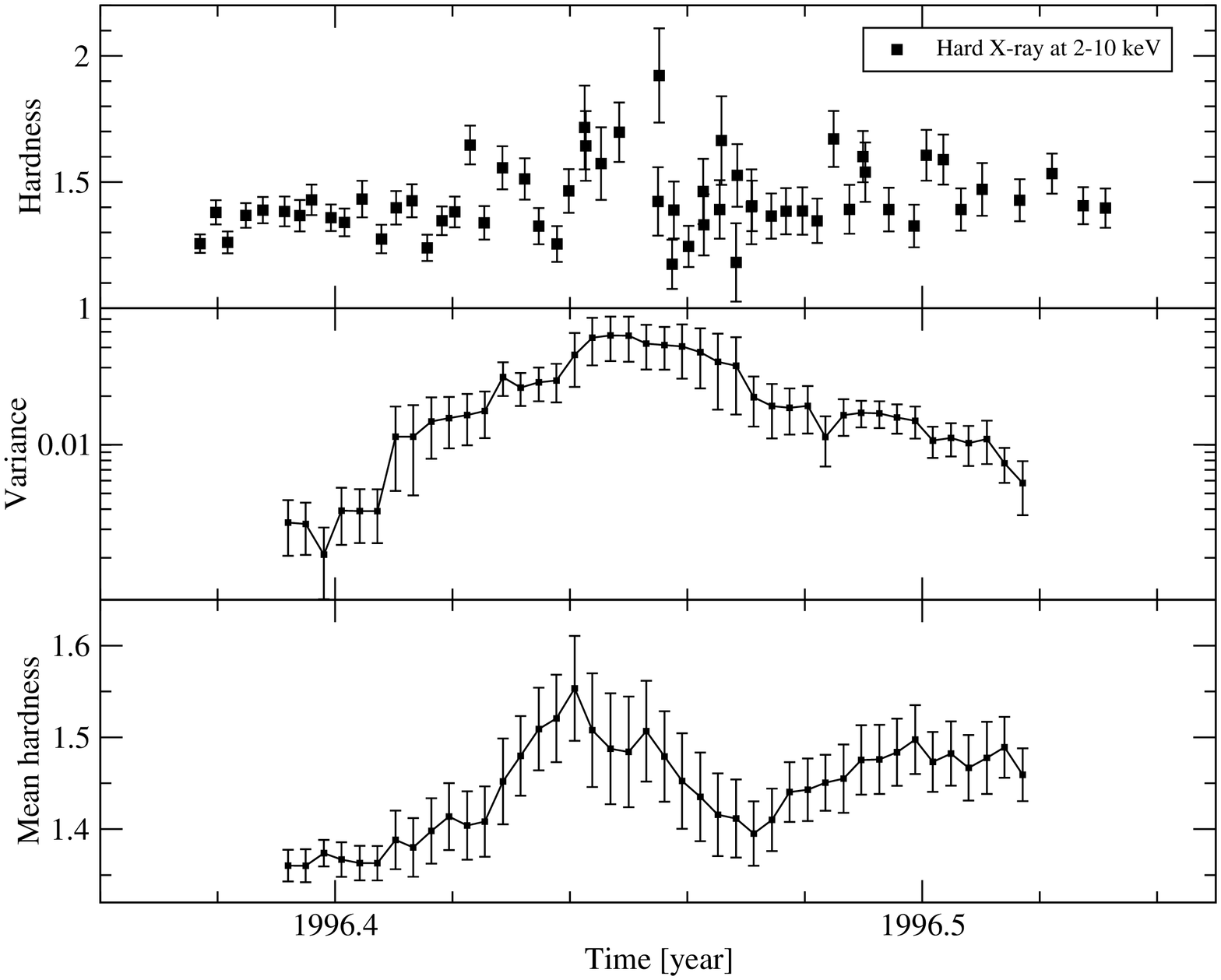}
\caption{ \emph{Left panels}. The epoch of ejection of the component
  C8 (red vertical line) superimposed onto the changes of the
  \emph{RXTE} light curve (at 2-20 keV \cite{gliozzi06}) and its
  variance (top panel), and the hardness ratio (defined as
  (15-6)\,keV/(6-2)\,keV), its variance and mean value (bottom panel)
  for a time period of about two years. \emph{Right panels}. The daily
  changes of the \emph{RXTE} light curve (at 2-10 keV
  \cite{gliozzi03}) and its variance (top panel), and hardness ratio
  ((10-6)\,keV/(6-2)\,keV), its variance and mean value (bottom panel)
  during a period of nearly two months. Note that component C7 was
  ejected in 1996.64 (see Fig.~2), approximately two months after the
  minimum in the X-ray light curve and is not included here.}
\label{fig5}
\end{figure}

The disk-jet coupling in 3C\,390.3 is further evidenced from the hard
X-ray monitoring data. In Fig.~\ref{fig5}, the ejection time of the
component C8 is compared with various characteristics of the X-ray
light-curve at 2-20 keV \cite{gliozzi06}. The jet ejection epoch
coincides with the dips in the intensity of the X-ray flux and of the flux
variability (left upper panel), the hardening of the spectrum (similar to
3C\,120) and a large variance of the hardness (left lower
panel). Similar characteristics such as minimized flux variability,
increased hardness and variance of the hardness are observed during
the dip in the X-ray light-curve at 2-10 keV (Fig.~\ref{fig5}, right
panels).

Similar to 3C\,120, the X-ray dip and the hardening of the spectrum are
followed by the ejection of a radio component in the 3C\,390.3. This
is a clear evidence that this correlation is a characteristic feature
for radio-loud galaxies (see also \cite{marscher06} in this
proceedings). In addition, we showed that during the X-ray dips the
variance of the hardness increases while the variance of flux
variability sharply becomes lower (Figs.~\ref{fig4},\ref{fig5}),
similar to what happens for the microquasar GRS\,1915+105
\cite{mirabel98}. This is a strong evidence for a common physical
mechanism of disk-accretion and jet production processes in
microquasars \cite{fender04} and AGN.


\section{Similarities and differences with microquasars}
AGN and microquasars share similar activities displaying superluminal
ejections of radio blobs from the central nucleus. The difference is
in the energy output and mass, suggesting that microquasars are
scaled-down versions of AGN. The scaling factor between 3C\,390.3 and
a typical microquasar is $\sim 10^{7}$ assuming a linear scaling
between the central BHs ($M_{\rm BH}\approx3\times10^{8}M_{\odot}$ for
3C\,390.3 and 10-15 $M_{\odot}$ for GRS\,1915+105). The X-ray
light-curves of AGN harboring a supermassive black holes can be used
to interpret small time scale variations occurring in microquasars,
which are not achievable by present observations. The duration of the
X-ray dip is about one to three months (see Figs.~\ref{fig4} and
\ref{fig5}, left panel) which translates to $\sim$(0.3 to 1) seconds
for a microquasar. Another useful parameter is the jet ejection rate
ranging from months to years in AGN \cite{kellermann04}. The rate of
ejections in 3C\,390.3 is about one component in two years
(Fig.~\ref{fig2}) which scales to a one ejection in every six seconds
in a microquasar. The high ejection rate of radio blobs in
microquasars is associated with X-ray dips occurring in the repeatable
transitions between the hard and soft states \cite{fender04}. In
GRS\,1915+105 the dips can take place on time scales from 1-2 seconds
(see \cite{soleri06} in this proceedings) to 200 seconds in the X rays
\cite{fender04}. The high dip rate (one dip every two seconds) is
comparable with the rate of expected ejections deduced from the mass
scaling. The low dip rate in GRS\,1915+105 translates to one ejection
in about hundred years in 3C\,390.3. If the analogy is correct, then the
ejection rate in 3C\,390.3 should vary from $\sim$(0.01 to 1)
ejections per year on time scales of thousand years.
We should bear in mind that the ejection of radio components in
radio-loud AGN occurs always in the hard state. State transitions
(hard to soft), at which the component is ejected in microquasars,
have not been observed in AGN so far.

The 3C\,390.3 shares also similar characteristics with the galactic
black hole candidate Cyg X-1 in the low/hard state. Gliozzi et
al. \cite{gliozzi06} argued that the simultaneous variations of
spectral and intrinsic timing properties in 3C\,390.3 are similar to
variations observed in Cyg X-1 and other galactic black holes. In the
same low/hard state, the ratio of timescales of the FWHM of X-ray
flares in 3C\,390.3 and Cyg X-1 is roughly consistent with the
dynamical mass scaling \cite{leighly97b} suggesting that the flare
time scale is proportional to the mass of the black hole.\\

This research has made use of data obtained from the High Energy
Astrophysics Science Archive Research Center (HEASARC), provided by
NASA's Goddard Space Flight Center.


\end{document}